\documentclass[mathleft
]{an}
\usepackage{graphicx}
\usepackage{times}
\begin{document}

\Pagespan{789}{}
\Yearpublication{2006}%
\Yearsubmission{2005}%
\Month{11}%
\Volume{999}%
\Issue{88}%

\title{XMM-Newton and Broad Iron Lines}

\author{A.C. Fabian\inst{1}\fnmsep\thanks{Corresponding author:
  \email{acf@ast.cam.ac.uk}\newline}
}
\titlerunning{}
\authorrunning{A.C Fabian}
\institute{Institute of Astronomy, Madingley Road, Cambridge CB3 0HA,
U.K.}

\received{}
\accepted{5}
\publonline{later}


\abstract{Iron line emission is common in the X-ray spectra of
  accreting black holes. When the line emission is broad or variable
  then it is likely to originate from close to the black hole. X-ray
  irradiation of the accretion flow by the power-law X-ray continuum
  produces the X-ray 'reflection' spectrum which includes the iron
  line. The shape and variability of the iron lines and reflection can
  be used as a diagnostic of the radius, velocity and nature of the
  flow. The inner radius of the dense flow corresponds to the
  innermost stable circular orbit and thus can be used to determine
  the spin of the black hole. Studies of broad iron lines and
  reflection spectra offer much promise for understanding how the
  inner parts of accretion flows (and outflows) around black holes
  operate. There remains great potential for XMM-Newton to continue to
  make significant progress in this work. The need for high quality
  spectra and thus for long exposure times is paramount.}

\maketitle

\section{Introduction}

Most of the radiation from luminous accreting black holes is released
within the innermost 20 gravitational radii (i.e. $20 r_{\rm
g}=20 GM/c^2$). In such an energetic environment, iron is a major
source of line emission, with strong lines in the 6.4--6.9~keV band
depending on ionization state. Observations of such line emission then
provides us with a diagnostic of the accretion flow, conditions and
the strong gravity in this regime (Fabian et al 2000; Reynolds \&
Nowak 2003; Matt 2006; Miller 2007). 

The rapid X-ray variability found in many Seyfert galaxies requires
that  the emission orginates at small radii. The high
frequency break in their power spectra, for example, corresponds to
orbital periods at $\sim20 r_{\rm g}$ and variability is seen at still
higher frequencies (Uttley \& McHardy 2004; Vaughan et al 2005). Key
evidence that the very innermost radii are involved comes from
Soltan's (1982) argument relating the energy density in radiation from
active galactic nuclei (AGN) to the local mean mass density in massive
black holes, which are presumed to have grown by accretion which
liberated that radiation. The agreement found between these quantities
requires that the radiative efficiency of accretion be 10 per cent or
more (Yu \& Tremaine 2002; Marconi et al 2005). This exceeds the 6 per
cent for accretion onto a non-spinning Schwarzschild black hole and
inevitably implies that most massive black holes are rapidly spinning
with accretion flows extending down to a just few $r_{\rm g}$.
Moreover, this is where most of the radiation in such accretion flows
originates.

\begin{figure}[]
\includegraphics[width=100mm]{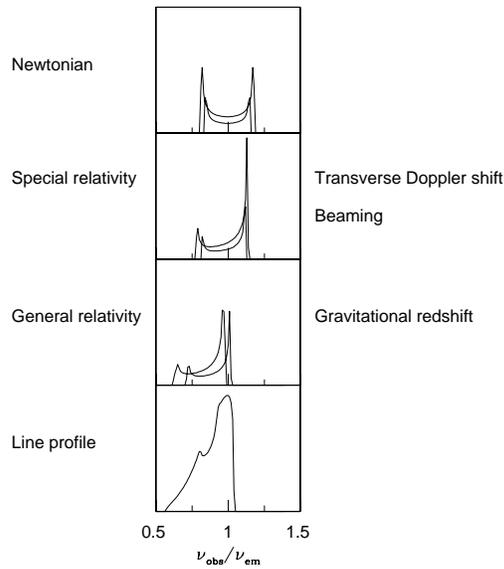}
\caption{The profile of an intrinsically narrow emission line is
  modified by the interplay of Doppler/gravitational energy shifts,
  relativistic beaming, and gravitational light bending occurring in
  the accretion disc (from Fabian et al 2000). The upper panel shows
  the symmetric double--peaked profile from two annuli on a
  non--relativistic Newtonian disc. In the second panel, the effects
  of transverse Doppler shifts (making the profiles redder) and of
  relativistic beaming (enhancing the blue peak with respect to the
  red) are included. In the third panel, gravitational redshift is
  turned on, shifting the overall profile to the red side and reducing
  the blue peak strength. The disc inclination fixes the maximum
  energy at which the line can still be seen, mainly because of the
  angular dependence of relativistic beaming and of gravitational
  light bending effects.  All these effects combined give rise to a
  broad, skewed line profile which is shown in the last panel, after
  integrating over the contributions from all the different annuli on
  the accretion disc. Detailed computations are given by Fabian et al
  (1989), Laor (1991), Dovciak et al (2004) and Beckwith \& Done
  (2004). }
\end{figure}

\begin{figure}[]
\includegraphics[height=80mm,angle=-90]{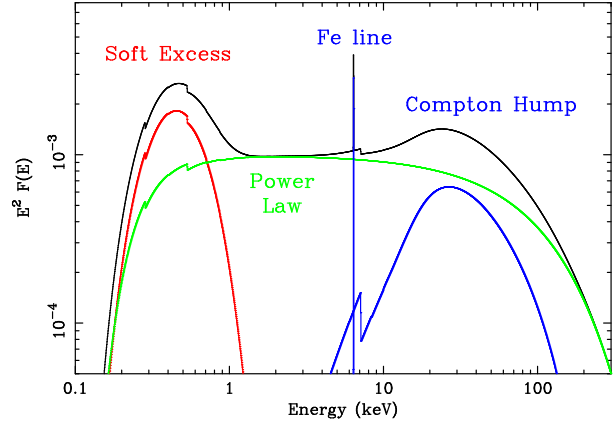}
\caption{ The
      main components of the X--ray spectra of an unobscured accreting BH
      are shown: soft X--ray emission from the
      accretion disc; power law from Comptonization of the soft
      X--rays in a corona above the disc; reflection continuum
      and narrow Fe line due to reflection of the hard X--ray emission
      from dense gas.  } 
\end{figure}

The X-ray spectra of AGN are characterized by several components: a
hard power-law which may turnover at a few hundred keV, a soft excess
and a reflection component (Fig.~1). This last component is produced
from surrounding material by irradiation by the power-law. It consists
of backscattered X-rays, fluorescence and other line photons,
bremsstrahlung and other continua from the irradiated surfaces.
Examples of reflection spectra from photoionized slabs are shown in
Fig.~2. At moderate ionization parameters ($\xi=F/n\sim
100$~erg~cm~s$^{-1}$, where $F$ is the ionizing flux and $n$ the
density of the surface) the main components of the reflection spectrum
are the Compton hump peaking at $\sim 30$~keV, the iron line at
6.4-6.9~keV (depending on ionization state) and a collection of lines
and reradiated continuum below 1~keV. When such a
spectrum is produced from the innermost parts of an accretion disk
around a spinning black hole then the outside observer sees it smeared
and redshifted (Fig.~3) due to doppler and gravitational redshifts
into the soft excess, the broad iron line and the Compton hump.

\begin{figure}[]
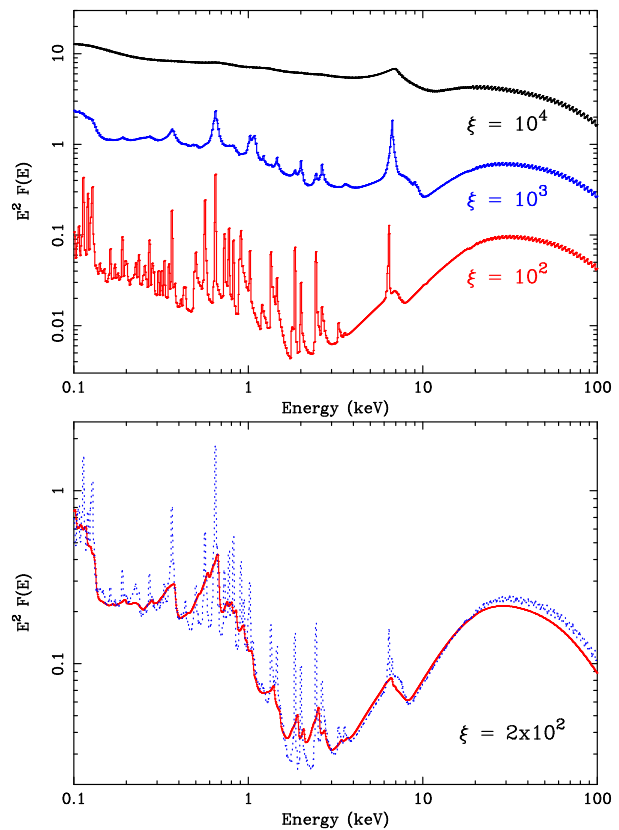

\includegraphics[height=80mm,angle=-90]{rossreflion.ps}
\includegraphics[height=80mm,angle=-90]{refblur.ps} 
\caption{{Upper panel}: Computed X--ray reflection spectra as a
  function of the ionization parameter $\xi$ ( Ross \& Fabian 2005).
  The illuminating continuum has a photon index of $\Gamma=2$ and the
  reflector is assumed to have cosmic (solar) abundances. {\it Lower
    panel}: Relativistic effects on the observed X--ray reflection
  spectrum (solid line).  We assume that the intrinsic rest--frame
  spectrum (dotted) is emitted in an accretion disc and suffers all
  the relativistic effects shown in Fig.~3. }
\end{figure}

All three main parts of the reflection spectrum have now been seen
from some AGN and Galactic Black Holes (GBH). The broad iron line and
reflection hump are clearly seen in the Seyfert galaxy MCG--6-30-15
and in the GBH J1650-400 (Fig.~4). More recently it has been realised
that the soft excess in many AGN can be well explained by smeared
reflection (Crummy et al 2005). 

\begin{figure}[]
\includegraphics[height=80mm,angle=-90]{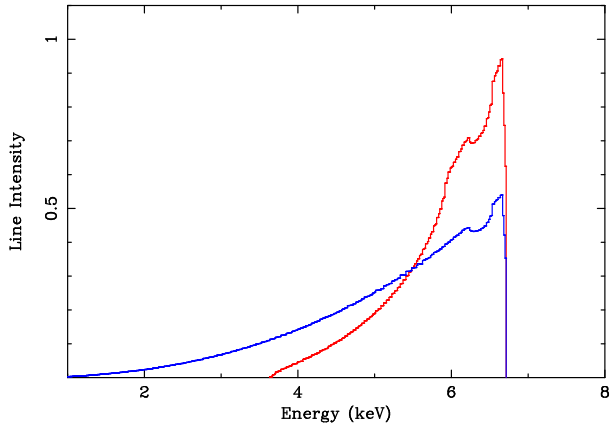} 
\caption{ The line profile dependence on the inner disc radius is
  shown for the two extreme cases of a Schwarzschild BH (red, with
  inner disc radius at $6~r_g$) and of a Maximal Kerr BH (blue, with
  inner disc radius at $\simeq 1.24~r_g$) }
\end{figure}
\begin{figure}[]
\includegraphics[height=80mm,angle=-90]{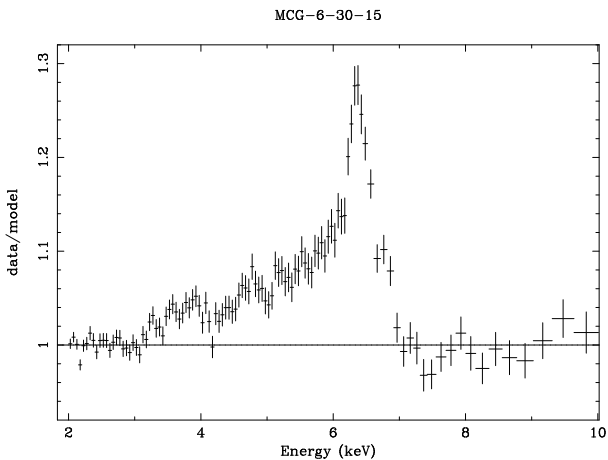} 
\caption{ The
  broad iron line in MCG--6-30-15 from the XMM observation in 2001 (Fabian
  et al.\ 2002a) is shown as a ratio to the continuum model. }
\end{figure}

The extent of the blurring of the reflection spectrum is determined by
the innermost radius of the disk (Fig.~4). Assuming that this is the
radius of marginal stability then enables the spin parameter $a$ of
the hole to be measured. Objects with a very broad iron line like
MCG--6-30-15 are inferred to have high spin $a>0.95$ (Dabrowski et al
1997, Brennemann et al 2006). Some (Krolik \& Hawley 2002) have argued that
magnetic fields in the disk can blur the separation between innermost
edge of the disk and the inner plunge region so that the above
assumption is invalid. This probably makes little difference
however since the reflection comes from very dense matter. Matter in
the plunge region very rapidly drops to a low density (Reynolds \&
Begelman 1997) and only very strong fields, much larger than are
inferred in disks, can stop this steep decline in density.

\section{Observations}

Broad iron lines and reflection components are seen in both AGN (e.g.
Nandra et al 1997) and GBH (Miller et al 2002, 2003, 2004) with
examples shown in Figs.~5,~6 and 7. They are not found in all objects or
in all accretion states. There are many possible reasons for this,
including overionization of the surface, low iron abundance, and
beaming of the primary power-law away from the disk (e.g. if the
power-law originates from the mildly-relativistic base of the jet).

\begin{figure}[]
\includegraphics[height=80mm,angle=-90]{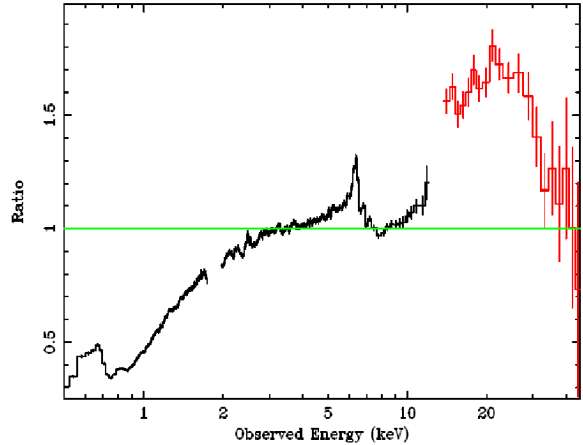} 
\caption{ Ratio of Suzaku data of MCG--6-30-15 to a power-law
  spectrum, shoing the broad iron line and the Compton hump (Miniutti
  et al 2007). The low
  energy drop is due to the warm absorber in the source. }
\end{figure}

\begin{figure}[]
\includegraphics[height=80mm,angle=-90]{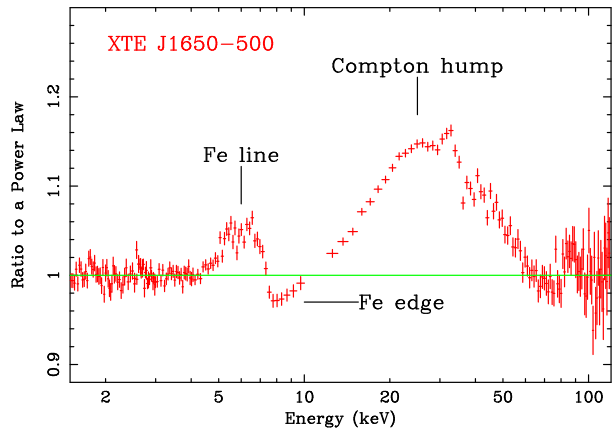}
\caption{Broadband BeppoSAX spectrum of XTE~J1650--500 (as a ratio to the
    continuum). The signatures of relativistically--blurred reflection
    are clearly seen (Miniutti et al 2004).}
\end{figure}
\begin{figure}[]
\includegraphics[height=80mm,angle=-90]{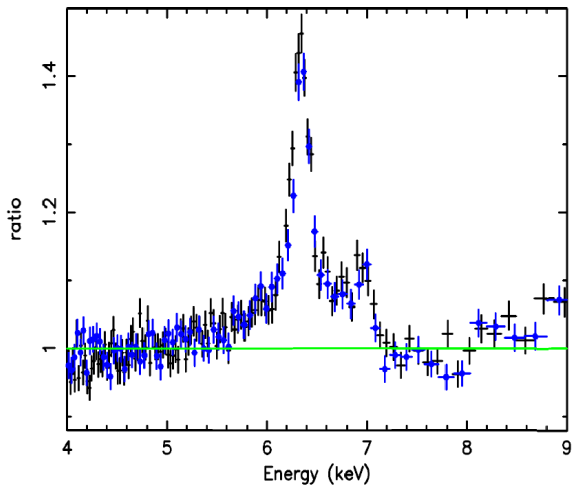} 
\caption{Suzaku spectrum (ratio to a power-law) shoing the iron-K
  region. Note the intense narrow iron line at 6.4~keV, presumably
  from distant matter, and the broad underlying iron line from the
  inner disc (Reeves et al 2007).}
\end{figure}

It should be noted that in many cases there will be a narrow iron line
component due to reflection from distant matter (Fig.~8 shows an
example where the narrow line dominates). Absorption due to
intervening gas, warm absorbers and outflows from the AGN, as well as
the interstellar medium in both our and the host galaxy must be
accounted for. Moreover, if most of the emission emerges from within a
few gravitational radii and the abundance is not high, then the
extreme blurring can render the blurred reflection undetectable
(Fabian \& Miniutti 2005). 

\begin{figure}[]
\includegraphics[height=80mm,angle=-90]{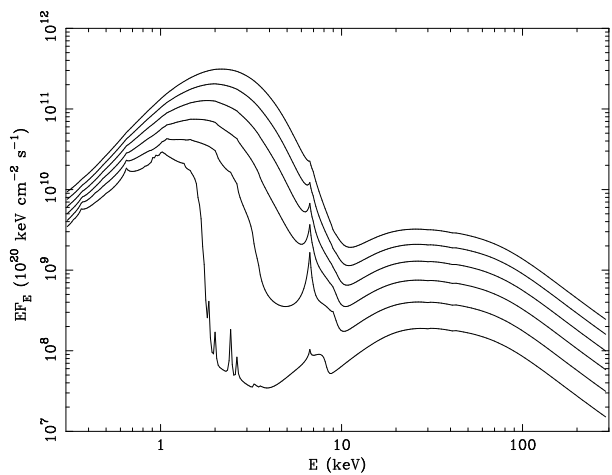} 
\caption{Irradiated disc spectra for accreting stellar mass black
  holes (Ross \& Fabian 2007). Thermal disc emission of increasing
  temperature from within the disc accounts for the large soft X-ray
  quasi-blackbody. The disc material is significantly hotter than for
  the AGN case meaning the the features undergo some Compton
  blurring. }
\end{figure}
\begin{figure}[]
\includegraphics[height=80mm,angle=-90]{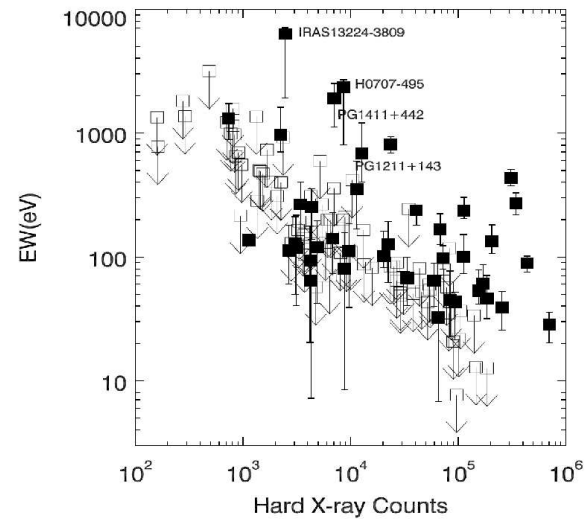} 
\caption{Results of the analysis of XMM AGN spectra by Guainazzi et al
(2006). Spectra with a broad line are shown with a filled square. More
than 100,000 cts are required to clearly find a broad line in many objects.}
\end{figure}

\begin{figure}[]
\includegraphics[height=80mm,angle=-90]{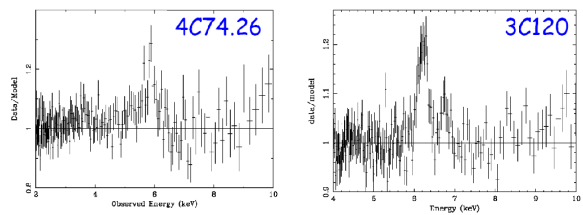} 
\caption{XMM spectra of the iron line in the radio-loud quasar
  4C\,74.26 (Ballantyne et al 2006) and the broad line radio galaxy
  3C\,120 (Ballantyne et al 2005). See Ballantyne (2007) for a
  discussion.  }
\end{figure}

In order to distinguish between the various spectral components, both
emission and absorption, we can use higher spectral resolution,
broader bandwidth and variability. An example of the use of higher
spectral resolution is the work of Young et al (2005) with the Chandra
high energy gratings. Observations of MCG--6-30-15 fail to show
absorption lines or feature associated with iron of intermediate
ionization. Such gas could cause some curvature of the appearent
continuum mimicking a very broad line. A broader bandwidth is very
useful in determining the slope of the underlying continuum. Some work
on this has been done with BeppoSAX (e.g. Guainazzi et al 1998) and
Suzaku (Miniutti et al 2007; Reeves et al 2007; Markowitz et al 2007).

Broad iron lines offer the possibility for studying the effect of jet
acceleration and powering on the accretion disc and flow. Some broad
lines have been seen from powerful radio sources (e.g. Fig.~11;
Ballantyne 2007), although much deeper exposures are required. Indeed
Guainazzi et al (2006) have shown that at least 150,000 counts are
needed in a typical AGN spectrum to clearly see the presence of a
broad line (Fig.~10). More than 50 per cent of XMM spectra of AGN
meeting that criterion do have broad lines. Nandra et al (2006, 2007)
have shown that the incidence of broad iron lines in XMM AGN spectra
is high, similar to that found with ASCA 10 years ago. Reports of a
lack of broad iron lines in the first year or two of XMM operations
(2000--2002) were presumably due to the relatively short exposures
adopted at that time, particularly in the guaranteed observing time.

\section{Variability}

A range of variability of the iron line is seen. In some objects (e.g.
NGC3516 Fig.~18, Iwasawa et al 2004; Mrk 766 Turner et al 2005) a
rather narrow component is seen to vary in energy, possibly in a
periodic manner. In others, redshifted and/or blueshifted compnents
come and go  (Dadina et al 2005; Turner et al 2004). In the
best objects where a very broad line is seen (e.g. MCG--6-30-15 Fabian
et al 2002; NGC4051 Ponti et al 2006) the reflection appears to change
little despite large variations in the continuum. The spectral
variability can be decomposed into a highly variable power-law and a
quasi-constant reflection component.  This behaviour is also borne out
by a difference (high-low) spectrum which is a power-law and the
spectrum of the intercept in flux-flux plots (Fig.~11). 

\begin{figure}[]
\includegraphics[height=80mm,angle=-90]{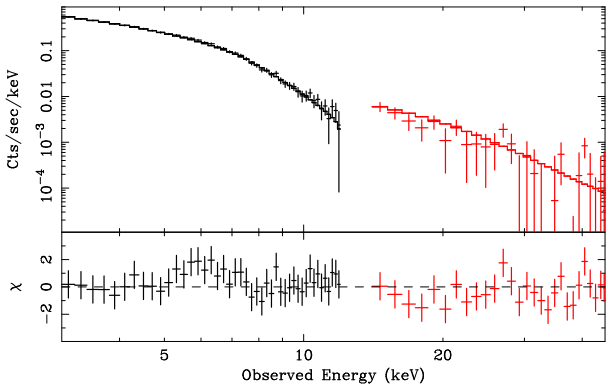} 
\caption{ Bright--faint difference spectrum for MCG--6-30-15 fitted
  with a power-law, showing that most of the spectral variability is due to the
power-law component changing in intensity (Miniutti et al 2007). }
\end{figure}

\begin{figure}[]
\includegraphics[height=80mm,angle=-90]{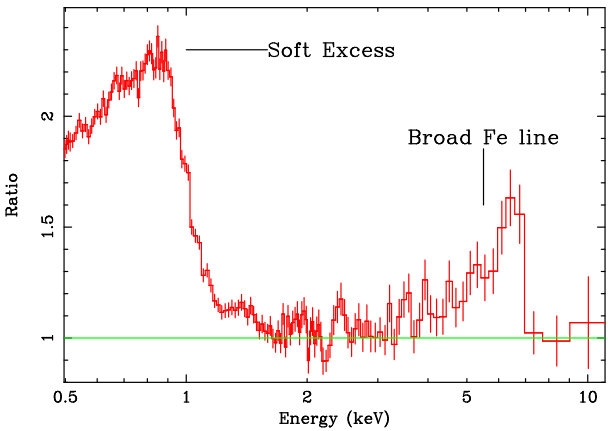} 
\caption{Ratio of the spectrum of the NLS1 galaxy 1H0707-494 to a
power-law. Spectral fits with either a very broad iron line or a
partial covering with a steep edge are equally good for this object
(Fabian et al 2004; Boller et al 2002).  }
\end{figure}

This behaviour was initially puzzling, until the effects of
gravitational light bending were included (Fabian \& Vaughan 2003;
Miniutti et al 2004; Miniutti \& Fabian 2005; Suebsuwong et al
2006). Recall that the extreme blurring in these objects means that
much of the reflection occurs within a few $r_{\rm g}$ of the horizon of
the black hole. The enormous spacetime curvature there means that
changes in the position of the primary power-law continuum have a
large effect on the flux seen by an outside observer (Fig.~13;
Martocchia \& Matt 1996). What this means is that an intrinsically
constant continuum source can appear to vary by large amounts just by
changing position in this region of extreme gravity. The reflection
component, however, will appear to be relatively constant in flux in
this region. Consequently, {\em the observed behaviour of these
  objects may just be a consequence of strong gravity.} Rapid motion
of the power-law source can, by special-relativistic beaming, also
introduce anisotropy in the irradiation (Reynolds \& Fabian 1997;
Beloborodov 1999) which contribute to the effects seen.

\begin{figure}[]
\includegraphics[height=80mm,angle=-90]{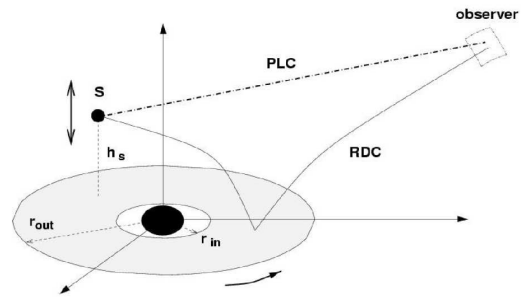} 
\caption{ Schematic of the light paths from source S at height $h$
  close to the black hole. The direct power-law component (PLC) and
  the reflection-dominated component (RDC) are indicated.   }
\end{figure}

\begin{figure}[]
\includegraphics[height=80mm,angle=-90]{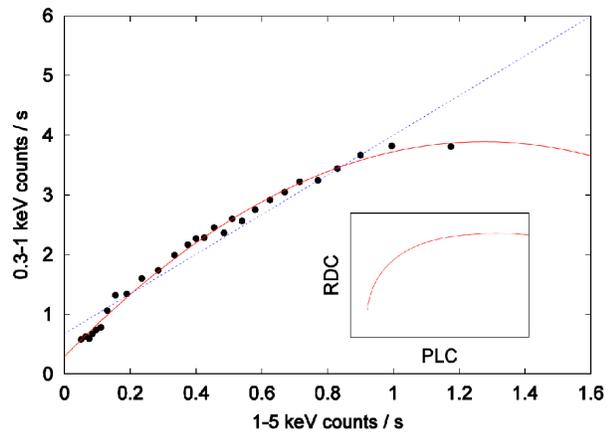} 
\caption{ Flux-flux plot for 1H\,0707. The 1--5~keV flux on the x-axis
is dominated by the PLC, whereas the 0.3--1~keV flux on the y-axis is
dominated by reflection (the RDC). The inset shows the 
behaviour expected from the light bending model.  }
\end{figure}
\begin{figure}[]
\includegraphics[height=80mm,angle=-90]{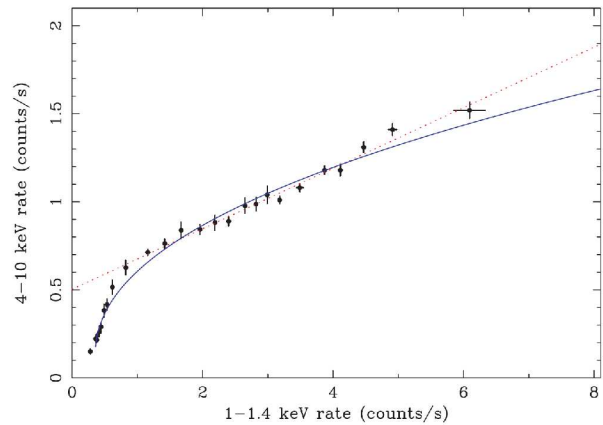} 
\caption{ Flux-flux plot for NGC\,4051 (Ponti et al 2006), analagous
  to Fig.~15. }
\end{figure}

Some of the Narrow-Line Seyfert 1 galaxies such as 1H0707 (Fig.~14) ,
IRAS13224 and 1H0439 appear to share this behaviour (Fabian et al
2002, 2004).  Some of these objects can show sharp drops around 7~keV
that may be interpreted as due to absorption from something only
partially covering the source. (If the covering was total then no
strong soft emission would be seen, contrary to observation.) However
the source is rapidly variable while being partially covered which
places what may be impossible constraints on what is doing the
covering. Also the GBH XTE\,J1650-500 behaved in a manner similar to
that expected from the light bending model (Rossi et al 2005).

The important point is that the X-ray spectra of many AGN have a
hard, blurred, reflection component which varies little in intensity
and is in shape consistent with reflection from the innermost part of
an accretion disc around a rapidly spinning black hole.

\section{Discussion}

Clear examples of relativistically-broadened iron lines are seen in
some AGN and GBH in some states. Such objects must have dense inner
accretion disks in order that the gas is not overionized. Detection of
a line is helped greatly if the iron abundance is super-Solar and if
there is little extra absorption due to very strong warm absorbers or
winds. Where broad lines are seen and can be modelled satisfactorally
then the spin of the central black hole can be reliably determined.

The study of absorption and emission variability of iron-K lines is in
its infancy, with some interesting and tantalising results produced so
far. The inner regions of accretion flows are bound to be structured
and so give rise to variations (e.g. Armitage \& Reynolds 2003). Some
may be due to motion or transience in the corona or primary power-law
source while others may reflect structure, e.g. spiral waves, on the
disk itself, intercepting primary radiation from much smaller radii.

\begin{figure}[]
\includegraphics[height=80mm,angle=-90]{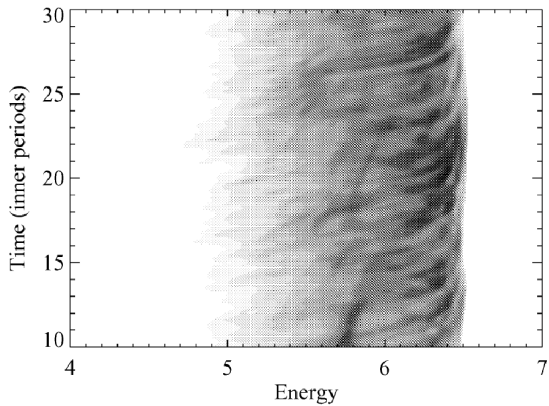} 
\caption{ Line variability expected from a turbulent accretion disc
  (Armitage \& Reynolds 2003).  }
\end{figure}
\begin{figure}[]
\includegraphics[width=80mm]{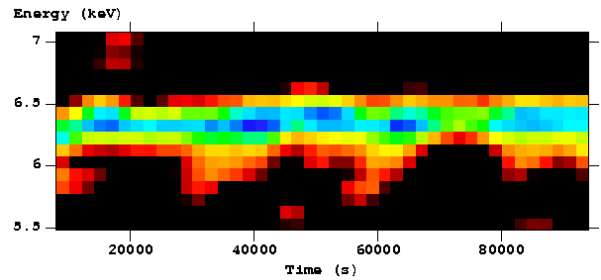} 
\caption{Iron line variability, suggestive of periodic behaviour, see
  in an XMM observation of NGC\,3516 (Iwasawa et al 2004).  }
\end{figure}

What XMM-Newton, and Suzaku, have established is the reality and
relative ubiquity of broad iron lines and other blurred reflection
features. We have a tool with which to study the inner workings of the
ultimate powerhouse in the Universe, the accreting black hole. To
examine variability on the light crossing time of the innermost orbits
of the black holes in the bright AGN is beyond XMM, since
square-metre-class collecting area is required, but there are enough
hints from the existing patchy observations for a) the problem to be
tackled in a statistical way, b) more black hole spins to be well
established, c) the inner regions of jetted sources to be explored,
and d) the position and/or motion of the power-law source to be
tracked. We need to map how the energy is extracted from the accretion
flow and deposited into the power-law generating corona, presumably by
magnetic fields. This is not a minor process but accounts for at least
12 per cent of the accretion power in high Eddington ratio sources and
50 per cent or so in the low Eddington ones ($L_{\rm Bol}/L_{\rm
  Edd}<0.1$; Vasudevan \& Fabian 2007).

This requires us to make much deeper exposures on many more suitable
AGN, to push them over the 150,000 ct boundary, an approach being
pursued by Matteo Guainazzi and colleagues. This does of course mean
long exposures on the next fainter level of sources. I favour very
long exposures on the brightest sources. Ms exposures on MCG--6-30-15,
for instance, or of the brightest NLS1s such as 1H0707 or
IRAS\,13224. We are only just beginning to explore these objects and
parameter space is ripe for serendipitous discovery as much as nailing
down the parameters we expect. As a community, we owe it to the legacy
of XMM-Newton to propose, accept and follow through such work which
will be invaluable for understanding the most powerful quasi-steady
sources in the Universe. It cannot be done any other way!

\acknowledgements I am grateful to many colleagues for collaboration and
discussion on broad iron lines, including Giovanni Miniutti, Jon
Miller, Chris Reynolds, Andy Young, Randy Ross, Luigi Gallo, Josefin
Larsson, Gabrielle Ponti and others. This is a heavily modified version of my
contribution for the XMM meeting on Broad Iron Lines 2006.

\end{document}